\documentclass{epl}
\usepackage{amsfonts}
\usepackage{amsmath}
\usepackage{amssymb}
\usepackage{graphicx}

\providecommand\bcdot{\boldsymbol{\cdot}}
\providecommand\bnab{\mbox{\boldmath $\nabla$}}

\begin{document}

\title{Droplet rebound from rigid or strongly hydrophobic surfaces}
\author{A. Gopinath}

\institute{
  Division of Engineering and Applied Sciences, Harvard
University, Cambridge, MA 02138 }

\pacs{nn.mm.xx}{First pacs description} \pacs{nn.mm.xx}{Second pacs
description} \pacs{nn.mm.xx}{Third pacs description}


\date{\today}%
\maketitle

\begin{abstract}
A weakly deformable droplet impinging on a rigid surface rebounds if
the surface is intrinsically hydrophobic or if the gas film trapped
underneath the droplet is able to keep the interfaces from touching.
A simple, physically motivated model inspired by analysis of
droplets colliding with deformable interfaces is proposed in order
to investigate the dynamics of the rebound process and the effects
of gravity. The analysis yields estimates of the bounce time that
are in very good quantitative agreement with recent experimental
data (Okumura et. al., (2003)) and  provides significant improvement
over simple scaling results.
\end{abstract}

\section{Introduction}

The collision and rebound of droplets has been a subject of interest
and investigation for many decades. Impacting drops  undergo
energetic bounces if the pressure in the gas film separating the
drops deforms the drop surfaces sufficiently to transform the drop's
kinetic energy into deformation energy before contact occurs.
Collisions in a incompressible, continuum gas usually result in a
bounce with contact occurring due to inter-particle forces or
surface imperfections$^{1,2}$. Bouncing can also result when drops
impinge on rigid  super hydrophobic$^{3}$ surfaces or on rigid walls
heated to well above the Leidenfrost temperature$^{4,5}$. Very
recently, Okumura et. al.$^{3}$, presented experimental data for the
collision and rebound, in air,  of $400-1000$ $\mu$m water drops
from a super-hydrophobic surface. They found that the measured
bounce time was consistently larger than that predicted by simple
scaling. Furthermore, with the drop radius held fixed, the contact
time was seen to increase significantly as the velocity of impact
decreased. This was attributed to effects of finite drop size and
gravity.

In this letter, a physically motivated model is proposed to
investigate this scenario, specifically  the dependence of the
bounce time and deformation on droplet size and velocity. Use of the
methodology developed in similar problems involving droplets
colliding with deformable interfaces$^{1,2}$, indicates that {\em
even when gravity is absent} the bounce time for a liquid drop
increases as the impact velocity decreases. Consideration of gravity
induced effects suggests that the bounce time is modestly modified
relative to the zero-gravity value. The predicted results are in
excellent quantitative agreement with experimental data of Okumura
et. al.,$^{3}$ showing that both effects are needed to account
properly for the details of the bounce.

\section{Dimensionless analysis yields reduced parameter space}

Consider a droplet of radius $a$ comprised of liquid with density
$\rho_{d}$, viscosity $\mu_{d}$ and interfacial tension $\sigma$
colliding with a rigid flat surface at impact speed $U_{c}$. Let the
ambient gas be incompressible and continuum with density $\rho_{g}$,
pressure $p_{o}$ and viscosity $\mu_{g}$. Dimensional analysis then
indicates that the parameters determining the characteristics of the
rebound process are - (a) the Weber numbers based on drop and gas
properties, $W_{d} \equiv \rho_{d}U_{c}^{2}a \sigma$ and $W_{g}
\equiv \rho_{g} U_{c}^{2}a / \sigma$ that measure the relative
importance of inertial pressures and surface pressure, (b) the
capillary numbers based on drop and gas viscosities, $C_{d} \equiv
\mu_{d} U_{c}/ \sigma$ and $C_{d} \equiv \mu_{g} U_{c}/\sigma$, and
(c) the Reynolds numbers $R_{g} \equiv aU_{c}\rho_{g}/\mu_{g} $ and
$R_{d} \equiv a U_{c}\rho_{d}/\mu_{d}$ characterizing viscous
effects in the gas and inside the drop. The effects of gravity are
quantified by the Froude number, $F_{d} \equiv U_{c}^{2}/(ag)$.
Droplet inertia is quantified by the Stokes number $S \sim
O(W_{d}/C_{g})$,that also measures the effectiveness of gas
viscosity in dissipating the kinetic energy of the drop.

A large number of physically relevant collisions occur at conditions
satisfying $W_{d} \ll 1$, $S \gg 1$, $R_{d} \rightarrow 0$, $R_{g}
\leq 1$, $W^{1/2}_{d}/R_{d} \rightarrow 0$ and $C_{g} \ll 1$. In
such cases, viscous dissipation inside the drop and in the gas
outside may be neglected during the collision process as long as the
droplet rebounds$^{1}$. That is, the dissipation in the gas film has
a small impact on the bounce provided we are far from the
bounce-contact transition. The primary role of the film is then to
cushion the droplet from physical contact with the surface. We
expect the dynamics to primarily depend on just two parameters -
$W_{d}$ and $F_{d}$ and seek a theoretical description of the
approach and rebound in this special limit.

\begin{figure}
\onefigure{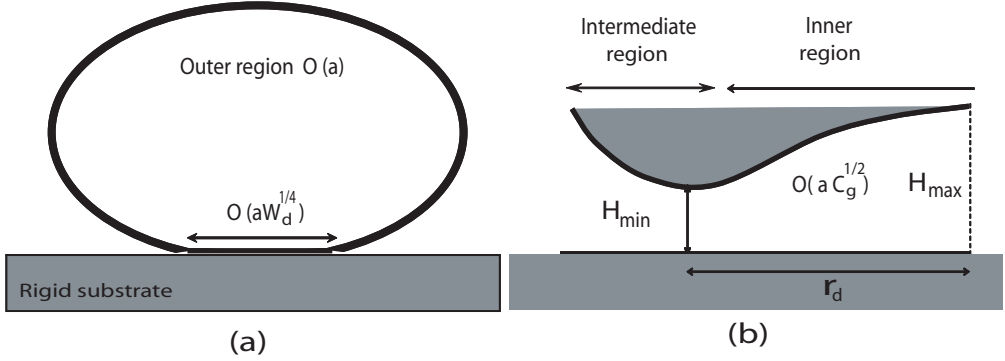} \caption{(a) Schematic of a drop of
initial radius $a$ approaching a rigid hydrophobic surface with
scaled velocity $W_{d}$. (b) A close up of the inner region
reveals a dimpled drop with a clearly defined rim. For a drop
supported by a film of gas, the gas pressure is $O(2\sigma/a)$.}
\end{figure}

\section{Collisions in zero gravity}

Consider for now collisions in gravity-less conditions so that
$F_{d}=\infty$. It is useful to first visualize the approach and
rebound scenario - for simplicity, let us choose a collision
occurring in the presence of a gas. For $S \gg 1$, changes in the
droplet velocity occur only when the gap thickness, $h_{o} \ll a$,
as a result of the large lubrication force in the gap. Drop
deformation becomes important when the gap thickness is
$O(aC^{1/2}_{g}) \ll a$. This is easily seen by balancing the
viscous force exerted by the gas with the capillary pressure
$O(2\sigma/a)$. The center of mass speed is chosen to be $U_c$ at
this point. As the gap becomes smaller, the confined air film leads
to a strong localized pressure that is $O(2\sigma/a)$, that when
seen from far seems to act over a temporally varying nearly flat
region of radial extent $r_{d}(t)$ as depicted in Figure (1a).
Closer examination of the near-contact region as depicted in Figure
(1b) reveals that the inner region may in fact have a more
complicated shape - for instance during the approach, a dimple like
shape is seen with a radius of roughly $r_{d}(t)$ and a gap
thickness that is a maximum at the centerline. Insofar as the bounce
process itself is concerned, the detailed shape of the inner region
is of secondary importance - what is important is the angular extent
of the {\em flat} region$^{1,2}$. As the radius of the flat region
increases, the center of mass moves slower as as more of the initial
kinetic energy is channeled into surface deformation modes. At some
point, the centers of mass of the drops reverse their direction of
motion and the drops begin to rebound - during this process the flat
region begins to shrink and finally vanishes. For a droplet
colliding with a super-hydrophobic surface we expect a similar
picture to apply. The reason being that due to the nature of the
surface-drop interaction, the contact angle is nearly $180^{o}$ and
the near-contact area is flat to leading order here as well. Thus,
drop is nearly spherical except at a small region where it is flat.
This shape is reminiscent of the shape found by Mahadevan and
Pomeau$^{8}$ for small drops rolling due to gravity.

Simple scalings$^{1,2,3}$ for the bounce time and deformation are
readily obtained. For the deformation energy to balance the $O(2 \pi
\rho_{d} a^{3} U^{2}_{c}/3)$ initial kinetic energy, the
characteristic magnitude of the interface deformation should scale
as $O(aW^{1/2}_{d}) \sim O(\rho_{d}U_{c}^{2} a^{3}/ \sigma)^{1/2}$.
The time scale for the deformation to relax is approximately
$O(aW_{d}^{1/2}U_{c}^{-1}) \sim O(\rho_{d} a^{3} / \sigma)^{1/2}$.
This value, which we denote by $\tau_{R}$, was first obtained by
Lord Rayleigh in his study of the vibrational modes of a drop$^{7}$.
Note that the Rayleigh time scale is independent of the impact
velocity. The angular extent of the flat region is obtained by a
force balance on the drop and is $O(aWe_{d}^{1/4})$.

To obtain a more accurate picture of the rebound process, the
equations for the flow inside the drop have to be solved. The
formulation follows directly from earlier work on collisions with
deformable interfaces$^{1,2}$. A spherical coordinate system that is
fixed in space and time in the interior of the droplet, such that
the origin lies at a distance $a$ from the rigid surface, is chosen.
The velocity field in the drop is divergence free and governed by
the Navier-Stokes equation. Scaling all lengths by $a$, time with
$aWe_{d}^{1/2}U_{c}^{-1}$, and the pressure by
$(\rho_{d}U_{c}^{2}W_{d}^{-1/2})$ and recognizing that the
$O(a^{2}\rho_{d}\mu_{d}^{-1})$ time scale for viscous diffusion of
momentum in the drop is much larger than the bounce time, we find
that the fluid in the drop undergoes inviscid impulsive motion
described by,
\begin{equation}
{\partial_{t}{{\bf u}_{d}}} = - \bnab p_{d}.
\end{equation}
Invoking the concept of a velocity potential, $\phi$ such that ${\bf
u}_{d} = \bnab \phi$, we find that $\nabla^{2}\phi =0$ and the
dimensional pressure field is given by
\[
 p'_{d} = p'_{o}+2{(\sigma/a)} - \rho_{d}{\partial_{t}
\phi'}
\]
$p'_{o}$ being the ambient pressure. The effect of the $O(2
\sigma/a)$ pressure field localized in a flat region of angular
extent $\alpha(t)$  is incorporated via the equation $ p'_{g} =
p'_{o} + 2 ({\sigma / a}) {\mathrm{H}}(\alpha(t)-\theta), $ $\theta$
being the angular co-ordinate of a cylindrical co-ordinate system as
in Figure (1a), and ${\mathrm{H}}$ being the Heaviside function. The
radial position vector of a surface point  at $\eta=\cos{\theta}$
and the velocity potential are expanded using the Legendre functions
${\mathcal{P}}_{k}$ of order $k$ and are given by
\begin{equation}
r^{*}_{s}(\eta,t) = a R(\eta,t) = a
(1+W_{d}^{1/2}\sum_{k=0}^{\infty}D_{k}(t){\mathcal{P}}_{k}(\eta))\:
\:\:\:\:{\mathrm{and}}\:\:\:\: \:\phi = \sum_{k=0}^{\infty} B_{k}(t)
(r^{*})^{k}{\mathcal{P}}_{k}.
\end{equation}
Application of the kinematic and normal stress boundary conditions,
the constant volume constraint along with simple geometry then
yields
\[
{d_{t} (B_{k})} = -{[{\mathcal{P}}_{k-1}(\cos{\alpha}) -
{\mathcal{P}}_{k+1}(\cos{\alpha})]} W_{d}^{-1/2}-[k(k+1)-2]D_{k},
\:\:\:\: \:\: {d_{t} (D_{k})} = kB_{k},
\]
\begin{equation}
{\mathrm{and }}\:\:\:\:\:
1+W_{d}^{1/2}\sum_{k=1}^{\infty}D_{k}(t){\mathcal{P}}_{k}(\cos{\alpha})
= (\cos{\alpha})^{-1}.
\end{equation}
Equation set (3) is then solved in conjunction with the initial
conditions $D_{k \geq 1}(t = 0) = 0$, $B_{k \geq 2}(t = 0) = 0$ and
$B_{1}(t = 0) = 1$. The rebound is complete when the center of mass
of the droplet reaches its initial position viz.,
$D_{1}(t=T_{b})=0$. Numerical solutions are obtained for various
values of $W_{d}$ by using a second-order predictor-corrector method
and varying the number of modes until convergent results are
obtained.

\begin{figure}
\onefigure{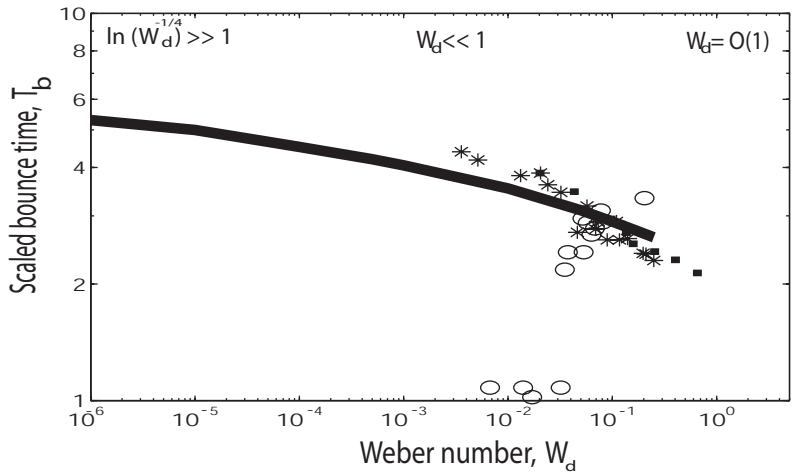} \caption{Scaled rebound time,
$T_{b} = T'_{b}/ \tau_{R} $ as a function of Weber number,
$W_{d}$. The solid curve is the theoretical prediction for  $S
\rightarrow  \infty$, $W_{d} \ll 1$ and $F_{d} = \infty $. The
symbols are experimental data points$^{3}$ for water drops in air.
The squares correspond to $a=0.6$ mm, the stars to $a=0.4$ mm and
the circles to $a= 1$ mm.}
\end{figure}

Figure (2) summarizes the results for $F_{d} = \infty$ and compares
them to experimental observations of Okumura et.al.$^{3}$. The solid
line corresponds to the theoretical prediction obtained by
numerically solving equation (3). We notice immediately that in
spite of the different impact velocities and sizes, there is a
reasonable collapse of the experimental data around the theoretical
prediction, indicating that $W_{d}$ is the relevant parameter to
use. Good qualitative and quantitative agreement is seen, especially
for the smaller size drops. Some of the data points for the 1 mm
drop do differ considerably from the predicted values but the
reasons for these are not clear. Nonetheless we find that the
theoretical result yields a good prediction even as the Weber number
becomes $O(0.1)$.

Both experimental data (that include gravity effects) and the
numerical results (without gravity) consistently indicate that as
$W_{d}$ decreases, the bounce time increases with decreasing
$U_{c}$. To investigate this behavior in more detail, analytical
asymptotes for the bounce time were obtained by a singular
perturbation analysis of (3). It is found that when
$\ln^{1/2}{(W_{d}^{-1/4})} \gg 1$, all the deformation modes do not
scale similarly as $O(aW^{1/2}_{d})$, instead a separation of scales
develops.  The $k=1$ mode corresponding to translation of the center
of mass scales as $O(aW_{d}^{1/2}\ln^{1/2}{(W_{d}^{-1/4})})$ while
the $k \geq 2$ surface deformation modes scale as
$O(aW_{d}^{1/2}\ln^{-1/2}{(W_{d}^{-1/4})})$. As a result, when
$W_{d} \rightarrow 0$, the highly localized pressure field causes
the $k \geq 2$ flow modes to become negligibly small. The radial
extent of the inner region then couples directly to the $k=1$
translational mode. In other words, rather than deform uniformly,
the approaching drop prefers to flatten the near contact region and
shift the center of mass appropriately in response to the ensuing
force. A spring like motion manifests itself wherein the center of
mass executes a harmonic oscillation$^{1,3}$  but like a non-linear
spring that {\em stiffens strongly} as $W_{d} \rightarrow 0$.In
fact, for asymptotically small $W_{d}$, the bounce time scales as,
\begin{equation}
T'_{b} \approx \pi \sqrt{(2/3)} \:[ \:a U_{c}^{-1}
W_{d}^{1/2}\ln^{1/2}{(W_{d}^{-1/4})}\:]
\end{equation}
while the maximum extent of the flat region scales as
$\alpha_{\mathrm{max}} = \:[\:{2 W_{d}}
/({3\ln{(W_{d}^{-1/4})}})\:]^{1/4}$.  As the Weber number increases,
all the deformation modes become equally important and the
separation of scales diminshes. The asymptotic expression in (4)
consistently under-predicts the actual bounce time since the
analytical approximation is obtained by ignoring the $k=2$ modes.
For finite $W_{d}$, the symmetry between approach and rebound is
broken. The effects of finite Weber number render the collision
process asymmetric, increasingly so as the Weber number approaches
$O(1)$ values.

\section{First effects of gravity}
The effects of gravity cannot be ignored when $W_{d} \sim
O(F_{d}^{2})$ i.e, when the impact velocity is $O(a^{3}\rho g^{2}/
\sigma)^{1/2}$ or smaller. To anticipate the effects of gravity we
consider the case of weak but non-zero gravity. That is, $F_{d}$ is
finite but $\varepsilon_{1} \equiv W_{d}^{1/2}F^{-1}_{d} \ll 1$.
Thus the droplet is still weakly deformed and the flow and
deformation modes are still described by (2). Note that the small
parameter $\varepsilon_{1}$ is the ratio of the critical speed at
which gravity becomes important and the impact speed - that is
$\varepsilon_{1} \equiv ag(a \rho_{d}/\sigma)^{1/2}U^{-1}_{c} \equiv
V_{g}/U_{c}$. A measure of the importance of gravity in affecting
the bounce time is then obtained by scaling the actual bounce time,
$T'_{b}$ with the gravity-free value $T^{0}_{b}$ and then plotting
this normalized rebound time as a function of a suitable small
parameter. Since the data for 400 micron drops  correspond most
closely to the limits we are investigating, these are chosen to test
the theoretical predictions.

Okumura et. al., proposed an approximate expression to incorporate
the effects of gravity which may be written as
\begin{equation}
({T'_{b} \over  T'^{0}_{b}})_{o} = 2(1 - {1 \over \pi} \cos^{-1}(
{\varepsilon_{1} \over {(1+ \varepsilon^{2}_{1})^{1/2}}})).
\end{equation}
In their expression, the zero-gravity bounce time, $(T^{0}_{b})_{o}
= (2.3 \tau_{R})$ is independent of $W_{d}$ and thus does not depend
on the impact velocity. The bounce time as predicted increases by at
most a factor of 2 compared to the zero-gravity value.  This
expression is plotted as a dashed curve in figure (3a) and compared
to the experimental obtained data points normalized the same way.
The circles indeed map onto to small $\varepsilon_{1}$ values but
the normalized bounce times are all much greater than unity. At
first sight this may seem to imply that gravity plays a crucial role
- however this is not quite correct. Since the true zero-gravity
bounce time is dependent on $W_{d}$, a more accurate reflection of
gravity effects is obtained upon incorporating this dependence.

Figure 3b illustrates precisely this effect. The squares are the
re-normalized data points obtained by calculating $T^{0}_{b}$
corresponding to the solid line in figure 2, and plotting
$T'_{b}/T^{0}_{b}$ as a function of the small parameter
$\varepsilon_{2} \equiv {
W_{d}^{1/2}\ln^{1/2}{(W_{d}^{-1/4})}F^{-1}_{d}}$. The data points
map to  low values of $\varepsilon_{2}$ as expected indicating the
weak effects of gravity. More importantly, the scaled bounce times
are within 20 \% of the gravity-free value suggesting that the
quantitative effects of gravity are relatively modest.

An approximate theoretical treatment allows us to anticipate the
dependence of the scaled bounce time on $\varepsilon_{2}$ observed
in figure 3b. For asymptotically small Weber number, the center of
mass of the drop behaves as if attached to a spring with a $W_{d}$
dependent stiffness. When the effects of an external gravity field
are included, an additional force acts on the center of mass due to
the weight of the drop. Set $r_{s}^{*}(\eta,t) = a (1 + \delta
D_{1}{\mathcal{P}}_{k}(\eta) + O(\delta_{2}))$ where $\delta_{1}
\equiv W_{d}^{1/2}\ln^{1/2}{(W_{d}^{-1/4})}$ and $\delta_{2} \equiv
W_{d}^{1/2}\ln^{-1/2}{(W_{d}^{-1/4})}$. The force balance on the
center of mass of the drop has the form
\[
(4 \pi \rho_{d}a^{3}/3)
(a\delta_{1})\:[\:U^{-1}_{c}(a\delta_{1})\:]^{-2}d^{2}_{t}(D_{1}) =
- 2 (\sigma /a) \pi (a\delta^{1/2}_{2})^{2} \alpha^{2} + (4 \pi
\rho_{d} a^{3}g/3)({\bf g}\bcdot{\bf U}_{c})\:(gU_{c})^{-1}.
\]
Algebraic manipulations using asymptotic properties of the Legendre
polynomials show that $\alpha^{2}(t) \approx D_{1}(t) +
O(\delta_{2})$.  The deformation mode $D_{1}$ then satisfies
\begin{equation}
d^{2}_{t}(D_{1}) = -{3 \over 2} D_{1} + {\delta_{1} \over
F_{d}}({\bf g} \bcdot {\bf U}_{c})(gU_{c})^{-1} = -{3 \over 2} D_{1}
+ \varepsilon_{2}
\end{equation}
the small parameter $\varepsilon_{2} \equiv {\delta_{1}/ F_{d}}$ and
gravity has been chosen to act in the direction of the initial
droplet motion. Equation (5) indicates that the relevant small
parameter quantifying effects of gravity when $W_{d} \ll 1$ is not
$\epsilon_{1}$ but $\epsilon_{2}$. Solving (5) with the initial
conditions $D_{1}(t=0)=0$ and $(d_{t}D_{1})(t=0)=1$ yields the
solution
\begin{equation}
D_{1}(t) = {2 \over 3}\varepsilon_{2} + \:[\:{2 \over 3}(1+{2 \over
3}\varepsilon_{2}^{2})\:]\:^{1 \over 2} \sin{(\sqrt{3 \over 2}t +
\tan^{-1}{(-\varepsilon_{2}\sqrt{2 \over 3})})}.
\end{equation}
This corresponds to the drop being spherical and moving with speed
$U_{c}$ when the collision process starts.

\begin{figure}
\onefigure{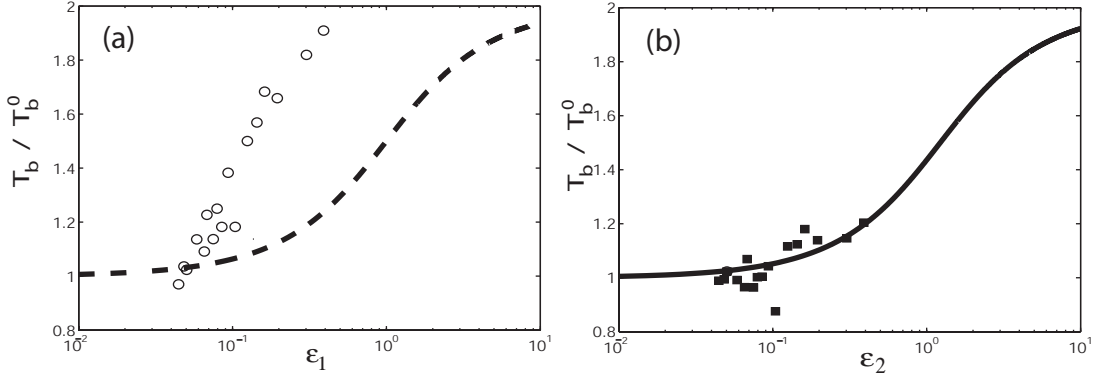} \caption{The effects of gravity on
the rebound time are explored in this plot. (a) Comparison of
equation (5) proposed by Okumura et. al. (dashed curve) with the
normalized bounce times (open circles) for 400 micron droplets
colliding in air. The x axis is $\varepsilon_{1} \equiv
V_{g}U^{-1}_{c}$. The gravity-free bounce time is independent of
$W_{d}$. (b) The solid line corresponds to equation (8) and
incorporates the effect of $W_{d}$ on the gravity-free bounce time.
The squares are the same data points re-normalized differently and
plotted as a function of $\varepsilon_{2} \equiv
V_{g}\ln^{1/2}{(W_{d}^{-1/4})}U^{-1}_{c}$.}
\end{figure}

It is easy to show from  equation (6) that when  $ 0 < \epsilon_{2}
\ll 1$, the modified bounce time is given by
\begin{equation}
{{T'_{b}} \over  {T'^{0}_{b}}} = ( 1 - {2 \over \pi}
\tan^{-1}{(-\varepsilon_{2}\sqrt{2 \over 3})}).
\end{equation}
It is clear that the bounce time increases relative to the
zero-gravity value by a factor that is between $1$ and $2$, the
latter value corresponding to $\epsilon_{2} \rightarrow \infty$.

Equation (8) shows  surprisingly good agreement with the
experimental data points which seems to suggest that gravity acts as
a weak perturbation for these droplets. Perhaps, a more careful
calculation in lieu of the simplified treatment just described will
result in a better prediction for larger values of
$\varepsilon_{2}$. Also note that the dashed and solid curves seem
to have the same form - this is of course primarily due to the way
we have chosen to normalize the curves. It should also be noted that
if the direction in which gravity acts is not the same as that of
the initial droplet motion, axi-symmetry is broken and the bounce
times can differ noticeably.

\section{Summary and conclusions}

To summarize, the collision and rebound dynamics of weakly deforming
drops impacting with velocities, $U_{c}$, satisfying $(a^{3}\rho
g^{2}/ \sigma) \ll {U_{c}}^{2} \ll ( a \rho /\sigma)^{-1}$, is
controlled by the interplay between the kinetic energy of the drop
and energy stored in surface deformation due to finite surface
tension, $\sigma$ with gravity acting as a weak perturbation. In the
limit of asymptotically small $W_{d}$, the scaled bounce time $T_{b}
= T'_{b}(\rho a^{3} / \sigma)^{-1/2}$, diverges as $
\ln^{1/2}\:[\:{(\rho U_{c}^{2}a /\sigma)^{-1/4}}\:]$ when $U_{c}
\rightarrow 0$. To leading order, gravity tends to increase the time
scale for the bounce relative to the zero-gravity value. Results of
numerical calculations indicate that the bounce time increases
relative to the asymptote as $U_{c}^{2} \rightarrow ( a \rho
/\sigma)^{-1}$ - a manifestation of the increasing effect of
non-linear deformation modes and of the drop being deformed
everywhere rather than merely locally.

Is the $O(aW^{1/4}_{d})$ local flattening seen in the $W_{d} \ll 1$
case, a precursor of the global deformation seen in pancake shaped
drops for $W_{d} \geq 1$? It is tempting to think that as $W_{d}
\rightarrow 1$, the maximum extent of the inner region grows till it
becomes comparable to the drop radius $a$ with the transition
between the inner and outer region becoming gentler. Consider
increasing the impact speed, $U_{c}$ such that it becomes $O(a \rho
/\sigma)^{-1/2}$ as happens when $W_{d} \sim O(1)$.  In this limit,
gravity plays a negligible role. The strongly deforming droplet
changes to a pancake like shape with characteristic radius $R_{d}
\geq a $ being different from the height $H_{d} \sim
O(a^{3}R^{-2}_{d})$. The initial kinetic energy is transferred
mainly into internal flow modes within the drop. The maximum extent
of the pancake drop is determined by a balance between the flow
driven by inertial acceleration and the flow due to the surface
tension induced pressure gradient at the edge of the pancake. This
yields the balance $(\rho_{d}U^{2}_{c}/a) \sim (\sigma
R^{4}_{d}/a^{6})$ thus yielding $(R_{d}/a)^{4} \sim
O(\rho_{d}U^{2}_{c} a /\sigma) \sim O(W_{d})$. Although the scalings
for the maximal {\em apparent contact area} seem to arise
differently for cases $W_{d} \ll 1$ and $W_{d} \gg 1$, it is
interesting and in some sense satisfying to note that they exhibit
the same dependence on $W_{d}$.

\acknowledgments I would like to thank Prof. L. Mahadevan for his
encouragement and help during the preparation of this manuscript.

\end{document}